\documentclass[9pt,twocolumn,twoside]{opticajnl}

\journal{opticajournal} 

\setboolean{shortarticle}{true}

\title{Lifshitz tail states in non-Hermitian disordered photonic lattices}

\author[1,2,*]{Stefano Longhi}

\affil[1]{Dipartimento di Fisica, Politecnico di Milano, Piazza L. da Vinci 32, I-20133 Milano, Italy}
\affil[2]{IFISC (UIB-CSIC), Instituto de Fisica Interdisciplinar y Sistemas Complejos - Palma de Mallorca, Spain}

\affil[*]{stefano.longhi@polimi.it}

\begin{abstract}
In lattices with uncorrelated on-site potential disorder, Anderson localization near the band edges can exhibit anomalously weak localization in the form of Lifshitz tail states. These states correspond to clusters of contiguous sites with nearly identical on-site energies, allowing excitations to extend significantly beyond the characteristic localization length determined by the inverse of Lyapunov exponent. Since Lifshitz tail states are rare events, with an exponentially small density of states, they are typically considered of limited practical importance. In this work, we demonstrate that when Anderson localization is induced by disorder in an imaginary on-site potential, Lifshitz tail states can dominate the system's dynamics and become experimentally observable. This phenomenon is illustrated through the Anderson-Bernoulli model in a non-Hermitian photonic lattice, shedding light on the unique interplay between disorder and non-Hermiticity in such systems.
\end{abstract}

\setboolean{displaycopyright}{false} 

\begin{document}

\maketitle

{\em Introduction.}  
Disorder-induced phenomena in wave systems, particularly Anderson localization  \cite{R1,R2,R3,R4,R5}, have attracted substantial interest across various fields, including condensed matter physics, optics, and acoustics. Anderson localization occurs when disorder is strong enough to disrupt wave transport, leading to the spatial confinement of eigenstates. This phenomenon has been observed in numerous physical systems, with photonics serving as a particularly captivating laboratory for investigating and observing such effects \cite{R6,R7,R8,R9,R10,R11,R12,R13,R14,R14b}. In one-dimensional lattices with uncorrelated on-site potential disorder, all eigenstates are exponentially localized, characterized by a localization length which is expected to be inversely related to the Lyapunov exponent. However, near the spectral band edges, certain eigenstates exhibit anomalously weak localization, forming the so-called Lifshitz tail states \cite{R4,R15,R16,R17,R18,R18b}. These rare states emerge from clusters of contiguous sites possessing nearly identical on-site energies, which facilitate significantly extended excitation compared to the typical localization length \cite{R4,R18}. This kind of "delocalization" at the tail can be detected by a marked deviation of the inverse participation ratio of the wave function from its Lyapunov exponent near the band edges \cite{R18}, a phenomenon very distinct from the well-established mobility edge
(localization-delocalization transition). Since Lifshitz tail states are extremely rare events and correspond to exponentially suppressed spectral densities \cite{R4}, they appear of limited relevance in conventional Hermitian systems and have largely been relegated to theoretical discussions \cite{R17,R18,R18b}.\\
In recent years, non-Hermitian systems --characterized by complex potentials-- have emerged as a fertile ground for novel wave phenomena, driven by their ability to model energy gain, loss, and nonreciprocal effects \cite{R19}. Non-Hermitian extensions of Anderson localization have uncovered phenomena distinct from their Hermitian counterparts, including the non-Hermitian delocalization transition \cite{R20,R21,R22,R23} and a form of universal non-Hermitian transport \cite{R24,R25,R26,R27,R28,R28b}. In these systems, wave propagation can occur in a jumpy manner even when the eigenstates of the Hamiltonian are exponentially localized \cite{R24,R28}. Lifshitz tail states are also present in non-Hermitian disordered lattices \cite{R29, R30}. Although their physical significance and practical relevance have remained uncertain \cite{R30}, recent studies \cite{R28,R28b} highlighted the role of the density of state tails in establishing the dynamical scaling in non-Hermitian transport.\\
In this Letter, we show that Lifshitz tail states, despite being extreme events, can dominate the {\em dynamical} behavior of disordered non-Hermitian lattices of {\em finite size}. Specifically, we focus on Anderson localization induced by imaginary on-site potential Bernoulli disorder in non-Hermitian photonic lattices. We demonstrate that, in such systems, Lifshitz tail states not only become more prominent but also play a pivotal role in shaping the overall system behavior. These findings challenge conventional views of Lifshitz tail states as negligible and highlight their experimental observability in non-Hermitian settings.
To illustrate this phenomenon, we analyze the Anderson-Bernoulli model, a paradigmatic framework for studying localization in a two-component alloy \cite{R17,R18b,R31,R32}, adapted here to a non-Hermitian photonic lattice. Our results provide insights into the dynamics of non-Hermitian systems with disorder and reveal potential avenues for harnessing non-Hermitian models to visualize rare events like Lifshitz tail states.\\
\\
{\em Model and Lifshitz tail states.} We consider a one-dimensional tight-binding lattice of size $L$ with a random on-site potential. The Schr\"odinger equation for a single particle hopping on the lattice reads
\begin{equation}
i \frac{d \psi_n}{dt}=J(\psi_{n+1}+\psi_n) +V_n \psi_n=\sum_{m=1}^{L} \mathcal{H}_{n,m} \psi_m
\end{equation}
where $\psi_n(t)$ is the wave amplitude at the $n$-th lattice site ($n=1,2,..,L$), $J$ is the hopping rate, $V_n$ is the on-site random potential, and $\psi_{0}(t)=\psi_{L+1}(t)=0$ for open boundary conditions. $V_n$ are assumed to be independent random variables with the same probability density distribution. For the Bernoulli-Anderson model \cite{R18b,R31,R32}, $V_n$ can take only two values, $V_n=W$ with probability $p$ and $V_n=0$ with probability $(1-p)$. Hence the on-site potential is a stochastic sequence of  0 and $W$.  In the  dissipative Anderson-Bernoulli model, the potential is imaginary, i.e. $W=-i \gamma$ where $\gamma>0$ is the particle loss rate. In this case, a photon occupying a site with $V_n=-i \gamma$ has a non-vanishing probability of being annihilated, thus providing a trap for photons \cite{R33}. Such a trapped configuration can be experimentally implemented in optical waveguide arrays with tailored losses or in dissipative synthetic lattices, such as in temporal mesh lattices \cite{R14b,R24,R34,R35}.
The eigenvalues $\lambda$ of the matrix Hamiltonian $\mathcal{H}$ are complex with ${\rm Im}(\lambda) \leq 0$, and we are interested in the behavior of $\lambda$ and corresponding eigenvectors $u_n$ in the large system size ($L \rightarrow \infty$) limit. For a large loss rate $\gamma \gg J$, the eigenvalues $\lambda$ and corresponding eigenvectors $u_n$ form two distinct groups, one with ${\rm Im}(\lambda) \simeq -\gamma$ corresponding to short-lived states, and the other one with  ${\rm Im}(\lambda) \simeq 0^-$ corresponding to long-lived states; see Fig.1(a). The eigenvectors of the former group have their excitations mainly in the lossy sites, where $V_n=-i\gamma$, while the latter group of eigenstates have their excitations mostly in the lossless sites, where $V_n=0$. The energy spectrum of both groups is symmetric about the imaginary axis ${\rm Re}(\lambda)=0$, which follows from the symmetry $V_n=-V_n^*$. As shown in the Supplemental document, this symmetry is however not relevant for the formation of Lifshitz tail states.\\ 
The dynamical evolution of the system is determined by the eigenstates possessing the longest lifetime \cite{R24,R28}, i.e. the eigenstates with ${\rm Im} (\lambda)$ closest to zero, which we call the Lifshitz tail eigenstates. They appear in pairs with  the same imaginary part of the eigenvalue but opposite real part. As shown in the inset of Fig.1(a), the 
 Lifshitz tail eigenstates with the longest lifetime are those with real part of energy $\lambda$ approaching the band edges $\pm 2J$ in the real part of the energy spectrum, and basically correspond to the low-energy eigenstates of the "particle-in-a-box" problem \cite{R18b,R32}, with the largest box length $l \sim L \rightarrow \infty$ as an extreme event. In fact, a realization of on-site potential $V_n$ can be viewed as random sequences of zeros (with stochastic length $l$) separated by sharp complex potential barriers $-i \gamma$ (the traps), i.e. stochastic sequences of potential wells of random length $l$ separated by infinite barriers in the $\gamma/J \gg 1$ limit. The Lifshitz tail eigenstates have their excitations mostly confined in the potential wells of large size $l$, for which the eigenstate $u_n$ can be approximated by the ground state of the infinite potential well, $u_n \simeq \sqrt{2/(l+1)} \sin [ n \pi/(l+1)] $[Fig.1(b)]. The corresponding eigenvalue $\lambda$ is close to the ground-state energy of the potential well and can be estimated from a perturbative analysis as (see the Supplemental document for technical details) 
 \begin{figure}[h]
 \centering
   \includegraphics[width=0.48\textwidth]{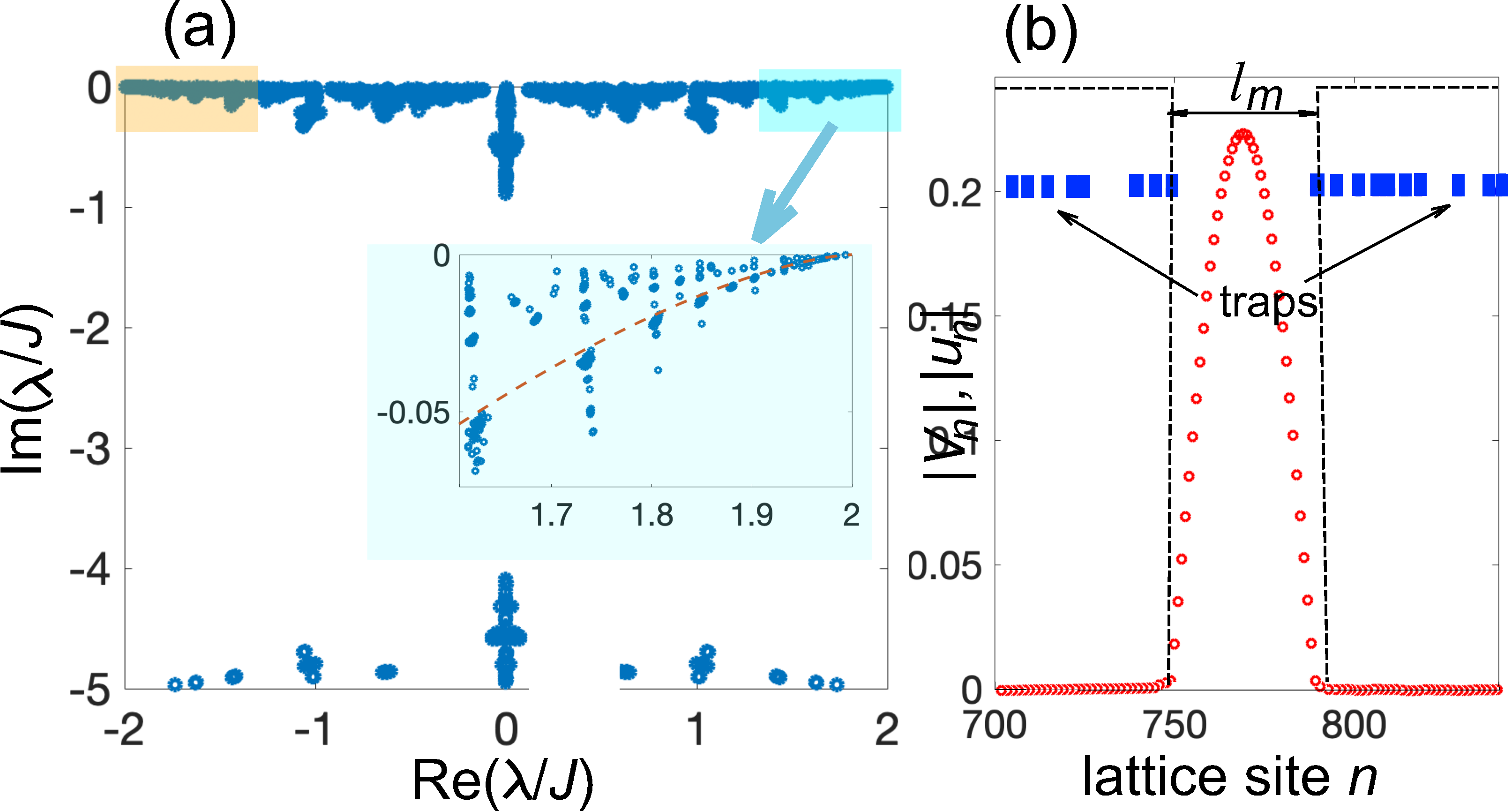}
   \caption{ \small (a) Behavior of the energy spectrum $\lambda$ of the matrix Hamiltonian $\mathcal{H}$ in complex plane for a given realization of the random Bernoulli potential $V_n$. Parameter values are $\gamma/J=5$, $p=0.2$ and $L=5000$. The spectrum is symmetry around the imaginary axis ${\rm{Re}} (\lambda)=0$ and can be grouped in two sets: the eigenvalues with small imaginary parts ($|{\rm Im}(\lambda)| \sim J^2/ \gamma$), corresponding to short-lived eigenstates, and the eigenvalues with large imaginary parts ($|{\rm Im}(\lambda)| \sim \gamma$), corresponding to long-lived eigenstates. The Lifshitz tail eigenstates are those with eigenvalues $\lambda$ close to the upper left and right corners, possessing extremely long lifetimes. The insets in (a) shows an enlargement of the spectrum near the upper right corner. The dashed curve in the panel depicts the loci of the eigenenergies of the Lifshitz tail states predicted by the "particle-in-a-box" model [Eq.(2)]. (b) Numerically-computed behavior of the Lifshitz tail state possessing the longest lifetime (modulus of $|u_n|$, red circles). The position of the traps is shown by filled blue rectangles. This eigenstate is well approximated by the ground state of a potential well of size $l_m$ (dashed rectangular curve), corresponding to the longest sequence with missing traps.}
 \end{figure}
 \begin{equation}
 \lambda=\lambda_r+i \lambda_i \simeq J \left(  2-\frac{\pi^2}{l^2}  \right)-i \left( \frac{1}{l^3}\frac{4 \pi^2 \gamma J^2}{\gamma^2+4J^2}    \right).
 \end{equation}
 Since the probability to find in the sequence a potential box of size $l$ is  $P_l=p(1-p)^l$, exponentially vanishing as $l \rightarrow \infty$, the Lifshitz tail states are rare, and in fact the densities of states $\rho(\lambda_{r})$ and $\rho(\lambda_{i})$
($ \lambda_{r,i}$ are the real and imaginary parts of $\lambda$) shows a typical exponential decay behavior at the band tails as in the Hermitian case \cite{R4,R18b,R32}. A major feature of Lifshitz tail states is that  their spatial extension, of order $\sim l$, is much larger and strongly deviates from the inverse of the Lyapunov exponent \cite{R18}: in fact, the latter remains finite at the band edges and describes the exponential decay tails of the wave function $u_n$ outside the potential well, while the potential box size $l$ can be as large as $\sim L$ in an extreme event. The "tail delocalization" effect  of Lifshitz states can be captured by computing the inverse participation ratio (IPR) \cite{R18} of the wave function $u_n$ in a box of length $l$ , which reads
\begin{equation}
{\rm IPR}_l=\frac{\sum_{n=1}^{L}{|u_n|^4}}{\sum_{n=1}^{L}{|u_n|^2}}=\frac{3}{2} \left( \frac{1}{l+1} \right).
\end{equation}
In a finite system of size $L$, the box length $l$ is clearly limited by $L$. Indicating by $l=l_m$ the largest potential well size in the lattice for a given realization of $V_n$, clearly $l_m$ is a stochastic variable with some probability distribution function. The mean value $\bar{l}_m$ of $l_m$ can be computed using the extreme value theory of stochastic Bernoulli processes and takes the approximate form (see e.g. \cite{proba})
\begin{equation}
\bar{l}_m \simeq \frac{\ln \left(\sigma pL \right)}{- \ln(1-p) }
\end{equation}
where $\sigma \sim 1$ is a form factor (we assume $1/ \sigma=\ln 2$ in the following analysis). Letting $l=\bar{l}_m$ in Eq.(3) and using Eq.(4), it follows that for Lifshitz tail states the IPR$_{l=\bar{l}_m}$ vanishes like $ \sim 1 / \ln (L)$ as $L \rightarrow \infty$. This vanishing law indicates some kind of delocalization, which is however weaker than both ergodic states, where the IPR scales as ${\rm IPR} \sim 1 /L$, and critical states, for which the IPR scales as ${\rm IPR} \sim 1/L^{D_2}$ with $0<D_2<1$ (fractal dimension). Our theoretical analysis of IPR scaling for Lifshitz tails is in excellent agreement with numerical simulations, as shown as an example in Fig.2(a).\\
\\
{\em System dynamics and Lifshitz tail delocalization.}   The delocalization features of the Lifshitz tail states are hard to be observed in large-size Hermitian systems, since they are extremely rare states and, for a given realization of the sequence $V_n$, their spatial locations in the lattice are unknown, rendering their excitation unfeasible. Conversely, in non-Hermitian lattices Lifshitz tail states not only become more prominent but also play a pivotal role in shaping the long-time system behavior. In fact, let us assume that at initial time {\em all sites} of the lattice are equally excited, i.e. let us assume $\psi_n(0)=(1/\sqrt{L})$. This condition ensures that Lifshitz tail states of the right edge band are surely excited, regardless of their unknown location and size in the lattice. The solution to the Schr\"odinger equation (1) reads
\begin{equation}
\psi_n(t)=\sum_{\alpha=1}^{L} C_{\alpha} u_n^{(\alpha)} \exp(-i \lambda_{\alpha} t)
\end{equation}
where $\lambda_{\alpha}$ and $u_n^{(\alpha)}$ are the $L$ eigenvalues and corresponding eigenvectors of $\mathcal{H}$, and $C_{\alpha}$ are the spectral excitation amplitudes, given by
$C_{\alpha}=\sum_{n=1}^L u_n^{(\alpha)}$, with $C_{\alpha} \simeq \sqrt{2/(1+l)} {\rm cotg} [ \pi/(2l+2) ] $ for a Lifshitz tail state of size $l$. Let us indicate by $\alpha=\alpha_0$ the index corresponding to the Lifshitz tail state with the longest lifetime, i.e. corresponding to the potential box of size $l=l_m$. Clearly, since all other states decay faster, in the long time limit $t \rightarrow \infty$ one has
\begin{equation}
\psi_n(t) \simeq C_{\alpha_0} u_n^{(\alpha_0)} \exp(-i \lambda_{\alpha_0} t)
\end{equation}
i.e. the long-time system's behavior is dominated by the {\em most rare} Lifshitz tail state. We note that, since the decay rate of Lifshitz states condense to zero as $L \rightarrow \infty$, a  longer propagation time will be required to reach the steady state as the system size $L$ is increased.
The emergence of such a state from the system's dynamics is a general phenomenon, observable for $W$ larger than $\sim J$, and can be detected by monitoring the temporal behavior of the {\em dynamical} IPR, which is defined as
\begin{equation}
{\rm IPR}(t)=\frac{\sum_{n=1}^L |\psi_n(t)|^4}{\sum_{n=1}^L |\psi_n(t)|^2}.
\end{equation}
According to the previous analysis, the statistical average of ${\rm IPR}(t)$, i.e. averaged over a large number of realizations of the Bernoulli disorder, should converge to the value ${\rm IPR}_{\bar{l}_m}$ as given by Eqs.(3) and (4), i.e. one has
\begin{equation}
 \lim_{t \rightarrow \infty } \overline { {\rm IPR}(t)}={\rm IPR}_{\bar{l}_m} \simeq \frac{3}{2} \frac{- \ln (1-p)}{\ln (\sigma pL)}.
\end{equation}
 \begin{figure}[h]
 \centering
   \includegraphics[width=0.48\textwidth]{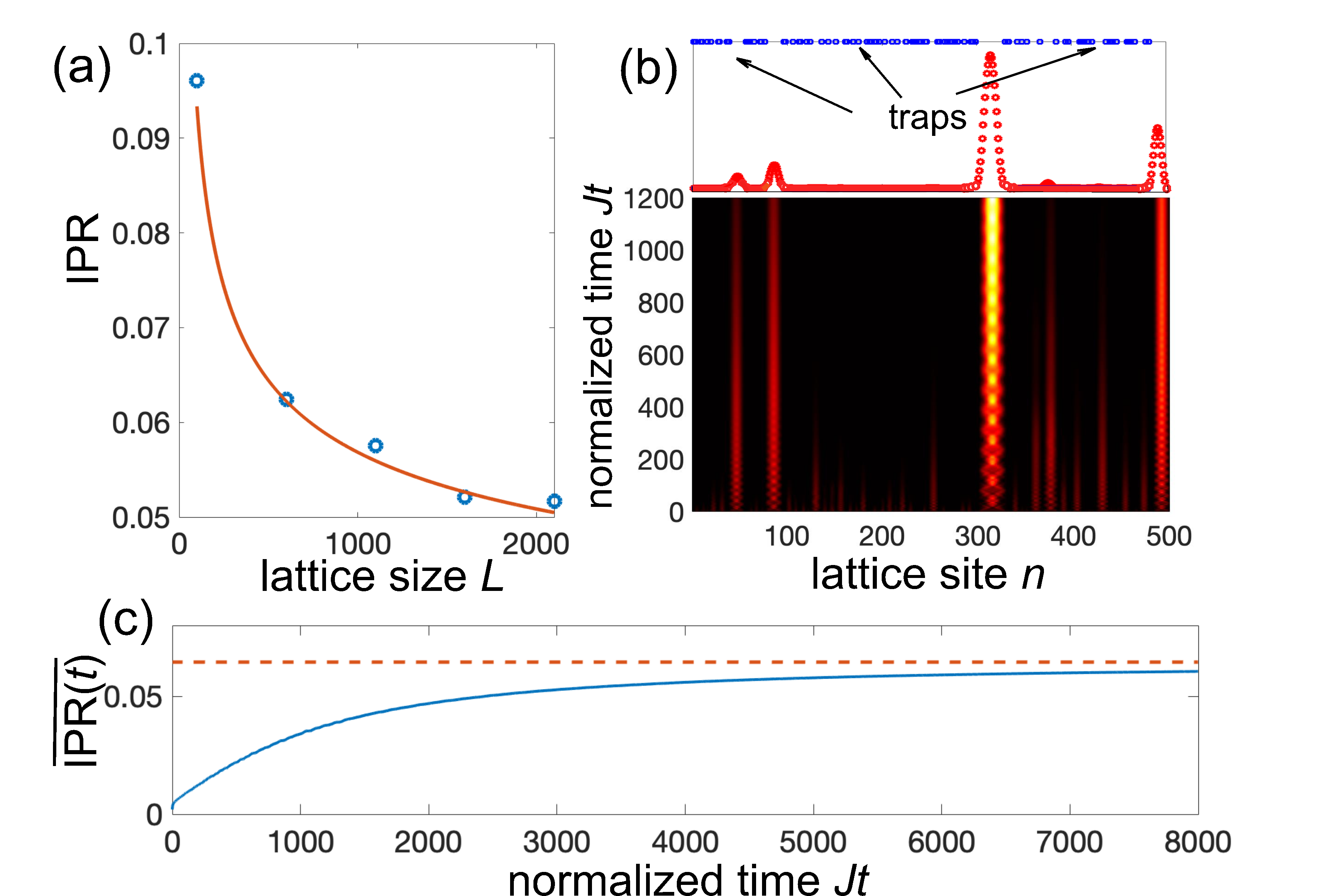}
   \caption{ \small (a) Numerically-computed behavior of the IPR of the Lifshitz tail state with the longest lifetime versus lattice size $L$ (circles), from $L=100$ to $L=2100$; the IPR is averaged over  200 realizations of Bernoulli disorder. Parameter values are $p=0.2$ and $\gamma/J=5$. The solid curve depicts the theoretically predicted behavior [Eqs.(3) and (4)]. (b) Dynamical evolution of $|\psi_n(t)|$ in a lattice of size $L=500$ for a single realization of Bernoulli disorder. Initial condition is $\psi_n(0)=1/\sqrt{L}$. The upper panel in (b) shows the behavior of $| \psi_n|$ at final propagation time $t=1200/J$ (red circles), and position of traps (blue squares). The main peak corresponds to the ground state of the potential box with largest size $l=l_m$. (c) Numerically-computed behavior of the dynamical IPR (solid curve), averaged over 100 realizations of disorder. The dashed horizontal curve in the asymptotic theoretical value predicted by Eq.(8).}
 \end{figure}
This theoretical prediction turns out to be in good agreement with the numerical results. As an example, Fig.2(b) shows the evolution of $|\psi_n(t)|$, normalized at each time step to its norm, in a lattice of size $L=500$ for a single realization of the Bernoulli disorder. At initial time, all sites of the lattice are equally excited.  As one can clearly see, in the long-time dynamics the system approaches the Lifshitz tail state with the longest lifetime, which is centered at the site $n=314$ and of size $l_m \simeq 28$, as shown in the upper panel of Fig.2(b). Figure 2(c) shows  the numerically-computed time evolution of the dynamical IPR, averaged over 100 different realizations of the Bernoulli disorder. Note that in the long time limit the mean value $ \overline{{\rm IPR}(t)}$ asymptotically approaches a constant value which is in good agreement with the theoretically predicted value given by Eq.(8).\\	
In the above analysis wee assumed 	a Bernoulli distribution of disorder, however Lifshitz tail states could be observed for other types of disorder, such as for a uniform imaginary disorder (see Supplemental document).\\
\\
 {\it Lifshitz tail states in synthetic temporal photonic lattices}.  An experimentally-accessible platform for the visualization of Lifshitz tail states is provided by photonic random walks in a synthetic temporal lattice with traps (see e.g. \cite{R14b,R24,R33,R34,R35}). The system  consists of two fiber loops of slightly different lengths $\mathcal{L} \pm \Delta \mathcal{L} $, that are connected by a variable directional coupler with a coupling angle $\beta$. Amplitude modulators are placed in the two loops, which introduce stochastic losses to emulate a Bernoulli distribution of  traps \cite{R33}. By discretizing the physical time as $t=t_n^{(m)}= mT +n\Delta T$, where $T= \mathcal{L} /c$ is mean travel time and $\Delta T=\Delta \mathcal{L} /c \ll T$ is the travel-time difference of light pulses in two loops, light dynamics is described by the set of discrete-time equations (see e.g. \cite{R24,R33})
 \begin{eqnarray}
u_n^{(m+1)} & = & \left( \cos \beta u_{n+1}^{(m)}+i \sin \beta w_{n}^{(m)} \right) \exp(- \gamma_n) \;\;\;\;\;\; \\
w_n^{(m+1)} & = & \left( i \sin \beta u_{n}^{(m)}+ \cos \beta w_{n-1}^{(m)} \right) \exp(- \gamma_n).
\end{eqnarray}
In the above equations, $u_n^{(m)}$ and $w_n^{(m)}$ are the pulse amplitudes at discrete time step $m$ and lattice site (unit cell) $n$ in the two fiber loops, $\beta$ is the coupling angle and $\gamma_n$ are uncorrelated stochastic losses (photonic traps) taking the two values $\gamma_n=\gamma>0$ with probability $p$ and  $\gamma_n=0$ with probability $(1-p)$. A finite lattice, containing $L$ unit cells, is obtained by assuming a coupling angle $\beta \neq \pi/2$ in the bulk, and $\beta=\pi/2$ at the two lattice edges. 
Like for the continuous random walk with traps [Eq.(1)], the long-time dynamics of the discrete random walk defined by Eqs.(9) and (10) is governed by the eigenstate of the propagator $\mathcal{U}=\exp(-i \mathcal{H})$ with the longest lifetime, which is localized at the potential box of the Bernoulli distribution with the largest size $l=l_m$. An example of such an eigenstate is shown in Fig.3(a). This state is  observed in the long-time random walk dynamics when one of the two fiber loops is seeded with a long optical pulse, such a flat-top or a Gaussian pulse, as shown in Fig.3(b) for the initial uniform condition $u_n^{(0)}=1/\sqrt{L}$, $w_n^{(0)}=0$ and in Fig.S4 of the Supplemental document for an input Gaussian excitation. 
Figure 3(c) shows the numerically-computed time evolution of the dynamical IPR, averaged over 200 realizations of disorder, clearly indicating stabilization to an asymptotic value which is in good agreement with the theoretical prediction [Eq.(8)].    
 \begin{figure}[h]
 \centering
   \includegraphics[width=0.46\textwidth]{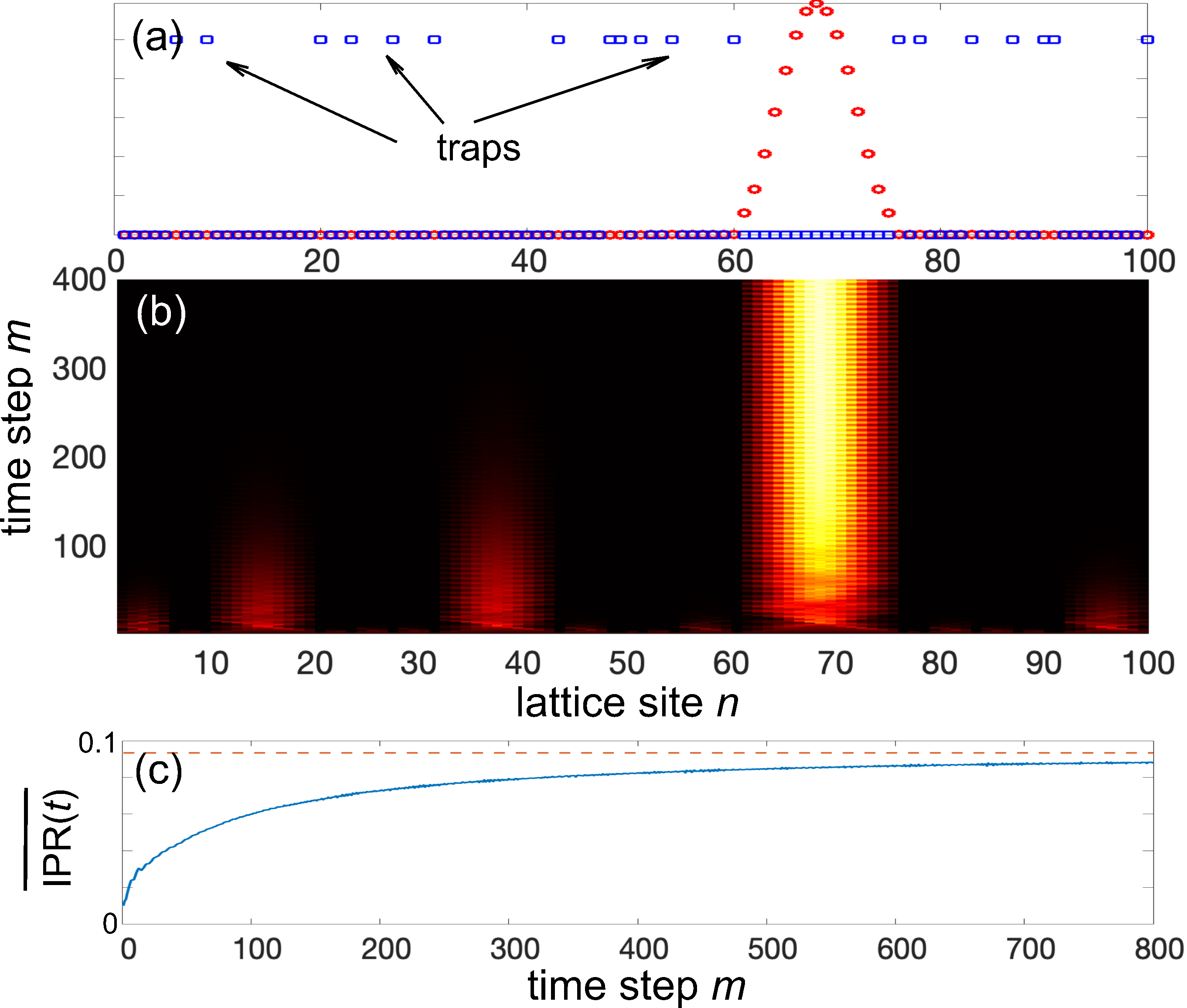}
   \caption{ \small (a) Numerically-computed behavior of the eigenstate (red circles) of the propagator $\mathcal{U}$ with the longest lifetime for the dissipative  photonic random walk. The position of the traps in the lattice are denoted by the blue squares. Parameter values are $\beta=\pi/6$, $L=100$, $p=0.2$ and $\gamma=2$. (b) Temporal light dynamics (behavior of $I_n=|u_n^{(m)}|^2+|w_n^{(m)}|^2$, normalized to its norm, versus propagation time step $m$ on a pseudocolor map) for uniform initial excitation $u_n^{(0)}=1/\sqrt{L}$, $w_n^{(0)}=0$. (c) Temporal evolution of the dynamical IPR versus time step $m$, averaged over 200 realizations of disorder. The IPR is defined as ${\rm IPR}=\sum_n I_n^2/ \sum_n I_n$. The dashed horizontal curve is the asymptotic value predicted by Eq.(8).}
 \end{figure}

{\em Conclusion.}
We have demonstrated that Lifshitz tail states, extremely rare states emerging in the theory of Anderson localization, become significantly more prominent in non-Hermitian systems. These states play a crucial role in shaping the dynamical behavior of finite-size lattices, as illustrated through the study of photonic random walks with traps. Our findings highlight the importance of Lifshitz tails in non-Hermitian dynamics and provide a pathway to experimentally visualize extreme events, such as rare weakly localized states, in photonic platforms. This work opens new directions for exploring the impact of disorder and non-Hermiticity in various wave-based systems.\\
\\ 
\noindent
{\bf Disclosures}. The author declares no conflicts of interest.\\
{\bf Data availability}. No data were generated or analyzed in the presented research.\\
\\
{\bf Funding}. Agencia Estatal de Investigacion (MDM-2017-0711).\\
{\bf Supplemental document}. See Supplement 1 for supporting content.

\newpage


 {\bf References with full titles}\\
 \\
 \noindent
1. P. W. Anderson, Absence of Diffusion in Certain Random Lattices, Phys. Rev. {\bf 109}, 1492 (1958).\\
2. D. J. Thouless, Electrons in disordered systems and the theory of localization, Phys. Rev. B {\bf 10}, 4134 (1974).\\
3. E. Abrahams, P. W. Anderson, D. C. Licciardello, and T. V. Ramakrishnan, Scaling Theory of Localization: Absence of Quantum Diffusion in Two Dimensions, Phys. Rev. Lett. {\bf 42}, 673 (1979).\\
4. B. Kramer and A. MacKinnon, Localization: theory and experiment, Rep. Prog. Phys. {\bf 56}, 1469 (1993).\\
5. F. Evers and A.D. Mirlin, Anderson transitions, Rev. Mod. Phys. {\bf 80}, 1355 (2008).\\
6. A. Lagendijk, B. van Tiggelen, and D.S. Wiersma, Fifty years of Anderson localization, Physics Today {\bf 62}, (8) 24 (2009).\\
7. D. S. Wiersma, P. Bartolini, A. Lagendijk, and R. Righini,  Localization of light in a disordered medium, Nature {\bf 390}, 671 (1997).\\
8. T. Schwartz, G. Bartal, S. Fishman, and M. Segev, Transport and Anderson localization in disordered two-dimensional photonic lattices, Nature {\bf 446}, 52 (2007).\\
9. Y. Lahini, A. Avidan, F. Pozzi, M. Sorel, R. Morandotti, D.N. Christodoulides, and Y. Silberberg, Anderson Localization and Nonlinearity in One-Dimensional Disordered Photonic Lattices, Phys. Rev. Lett. {\bf 100}, 013906 (2008).\\
10. A. Schreiber, K. N. Cassemiro, V. Potocek, A. Gabris, I. Jex, and Ch. Silberhorn, Decoherence and Disorder in QuantumWalks: From Ballistic Spread to Localization, Phys. Rev. Lett. {\bf 106}, 180403 (2011).\\
11. U. Naether, Y.V. Kartashov, V.A. Vysloukh, S. Nolte, A. T\"unnermann, L. Torner, and A. Szameit,
Observation of the gradual transition from one-dimensional to two-dimensional Anderson localization, Opt. Lett. {\bf 37}, 593 (2012).\\
12. S. St\"utzer, Y. V. Kartashov, V. A. Vysloukh, A. T\"unnermann, S. Nolte, M. Lewenstein, L. Torner, and A. Szameit, Anderson cross-localization, Opt. Lett. {\bf 37}, 1715 (2012).\\
13. S. Karbasi, C.R. Mirr, P.G. Yarandi, R.J. Frazier, K.W. Koch, and A. Mafi,
 Observation of transverse Anderson localization in an optical fiber, Opt. Lett. {\bf 37}, 2304 (2012).\\
14. M. Segev, Y. Silberberg, and D.N. Christodoulides, Anderson localization of light,
Nature Photon. {\bf 7}, 197 (2013).\\
15. I.D. Vatnik, A. Tikan, G. Onishchukov, D.V. Churkin, and A.A. Sukhorukov, Anderson localization in synthetic photonic lattices,
Sci. Rep. {\bf 7}, 4301 (2017).\\ 
16. I. M. Lifshitz, The energy spectrum of disordered systems, Adv. Phys. {\bf 13}, 483 (1964).\\
17. B.I. Halperin and M. Lax, Impurity-Band Tails in the High-Density Limit. I. Minimum Counting Methods,
Phys. Rev. {\bf 148}, 722 (1966).\\
18. G. Stolz, An introduction to the mathematics of Anderson localization, Contemp. Math {\bf 552}, 71 (2011).\\
19.  S. Johri and R. N. Bhatt, Singular Behavior of Eigenstates in Anderson's Model of Localization,
Phys. Rev. Lett. {\bf 109}, 076402  (2012).\\
20.  M. Bishop, Vi. Borovyk, and J. Wehr, Lifschitz Tails for Random Schr\"odinger Operator
in Bernoulli Distributed Potentials, J. Stat. Phys. {\bf 160}, 151 (2015).\\
21. Y. Ashida, Z. Gong, and M. Ueda, Non-Hermitian Physics, Adv. Phys. {\bf 69}, 3 (2020).\\
22. N. Hatano and D.R. Nelson, Localization Transitions in Non-Hermitian Quantum Mechanics,
Phys. Rev. Lett. {\bf 77}, 570 (1996).\\
23. P.W. Brouwer, P.G. Silvestrov, and C.W.J. Beenakker, Theory of directed localization in one dimension,
Phys. Rev. B {\bf 56}, R4333 (1997).\\
24. L.G. Molinari, Non-Hermitian spectra and Anderson localization, J. Phys. A: Math. Theor. {\bf 42}, 265204 (2009).\\
25. Q. Lin, T. Li, L. Xiao, K. Wang, W. Yi, and P. Xue,  Observation of non-Hermitian topological Anderson insulator in quantum dynamics,
Nature Commun. {\bf 13}, 3229 (2022).\\
26. S. Weidemann, M. Kremer, S. Longhi, and A. Szameit,
 Coexistence of dynamical delocalization and spectral localization through stochastic dissipation,
Nature Photon. {\bf 15}, 576 (2021).\\
27. A.F. Tzortzakakis, K.G. Makris, A. Szameit, and E.N. Economou,
Transport and spectral features in non-Hermitian open systems, Phys. Rev. Research {\bf 3}, 013208 (2021).\\
28. A. Leventis, K.G. Makris, and E.N. Economou, 
Non-Hermitian jumps in disordered lattices,
Phys. Rev. B {\bf 106}, 064205 (2022).\\
29. S. Longhi, Anderson localization in dissipative lattices, Ann. Phys. {\bf 535}, 2200658 (2023).\\
30. B. Li,  C. Chen, and Z. Wang, Universal non-Hermitian transport in disordered systems, arXiv:2411.19905 (2024).\\
31.  Z.-Y. Xing, S. Chen, and H. Hu, Universal Spreading Dynamics in Quasiperiodic Non-Hermitian Systems, arXiv:2412.01301 (2024).\\
32. A.V. Izyumov and B.D. Simons, Optimal Fluctuations and Tail States of Non-Hermitian Operators, Phys. Rev. Lett. {\bf 83}, 4373 (1999).\\
33. P. G. Silvestrov, Extended tail states in an imaginary random potential, Phys. Rev. B {\bf 64}, 075114 (2001).\\
34. R. Carmona, A. Klein, and F. Martinelli,  Anderson localization for Bernoulli and other singular potentials, Commun. Math. Phys. {\bf 108}, 41(1987).\\
35. F. Martinelli  and L. Micheli, On the Large-Coupling-Constant Behavior of the
Liapunov Exponent in a Binary Alloy, J. Stat. Phys. {\bf 48}, 1 (1987).\\
36. S. Longhi, Photonic random walks with traps, Opt. Lett. {\bf 49}, 2809 (2024).\\
37.  A. Regensburger, C. Bersch, B. Hinrichs, G. Onishchukov, A. Schreiber, C. Silberhorn, and U. Peschel, Photon propagation in a discrete fiber network: an interplay of coherence and losses, Phys. Rev. Lett. {\bf 107}, 233902 (2011).\\
38. M. Wimmer, M.-A. Miri, D. Christodoulides, and U. Peschel, Observation of Bloch oscillations in complex PT-symmetric photonic lattices, 
Sci. Rep- {\bf 5}, 17760 (2015).\\
39. L. Gordon, M.F. Schilling, and M.S. Waterman, An Extreme Value Theory for Long Head Runs, Probab. Th. Rel. Fields {\bf 72}, 279 (1986).

\end{document}